\documentclass[useAMS,usenatbib,fleqn]{mn2e}
\usepackage{amsmath,amssymb}
\usepackage{bm}
\usepackage{graphicx}

\title[Early Cosmic Merger of Multiple Black Holes]{Early Cosmic Merger of Multiple Black Holes}
\author[H. Tagawa, M. Umemura, N. Gouda, T. Yano \& Y. Yamai]{H. Tagawa$^{1,2}$\thanks{E-mail:
email@address ; tagawahr@nao.ac.jp }, M.Umemura$^{3}$, N. Gouda$^{1,2}$, T. Yano$^{2}$, Y. Yamai$^{4}$\\
$^{1}$The University of Tokyo, 7-3-1 Hongo Bunkyo, Tokyo 113-0033, Japan,\\
$^{2}$National Astronomical Observatory of Japan, 2-21-1 Osawa, Mitaka, Tokyo
181-8588, Japan,\\
$^{3}$The University of Tsukuba\\
$^{4}$Mitsubishi Space Software Co., Ltd.}
\begin{document}

\date{Accepted May 12, 2015}

\pagerange{\pageref{firstpage}--\pageref{lastpage}} \pubyear{2014}

\maketitle

\label{firstpage}

\begin{abstract}
We perform numerical simulations on the merger of multiple black holes (BHs) 
in primordial gas at early cosmic epochs. We consider two cases of BH mass: $M_\mathrm{BH}=30~M_\odot$
and $M_\mathrm{BH}=10^4~M_\odot$. 
Attention is concentrated on the effect of the dynamical friction by gas
in a host object. The simulations incorporate such general relativistic effects as
the pericentre shift and gravitational wave emission. 
As a result, we find that multiple BHs are able to merge into one BH 
within 100 Myr in a wide range of BH density. 
The merger mechanism is revealed to be categorized into three types: 
gas-drag-driven merger (type A), interplay-driven merger (type B), and 
three-body-driven merger (type C). 
We find the relation between the merger mechanism and the ratio of 
the gas mass within the initial BH orbit 
($M_\mathrm{gas}$) to the total BH mass ($\sum M_\mathrm{BH}$). 
Type A merger occurs if $M_\mathrm{gas}\gtrsim 10^5 \sum M_\mathrm{BH}$, 
type B  if $M_\mathrm{gas}\lesssim 10^5\sum M_\mathrm{BH}$, and 
type C if  $M_\mathrm{gas}\ll 10^5\sum M_\mathrm{BH}$.
Supposing the gas and BH density based on the recent numerical simulations on first stars,
all the BH remnants from first stars are likely to merge into one BH through the type B or C 
mechanism. 
Also, we find that multiple massive BHs ($M_\mathrm{BH}=10^4 ~M_\odot$) distributed over
several parsec can merge into one BH through the type B mechanism, 
if the gas density is higher than $5\times10^6~{\rm cm}^{-3}$. 
The present results imply that the BH merger may contribute significantly to 
the formation of supermassive BHs at high redshift epochs. 
\end{abstract}


\section{INTRODUCTION}
In the last two decades, it has been demonstrated that massive galaxies harbor 
supermassive black holes (SMBHs)
in the centres of their bulge components (Kormendy \& Ho 2013, and references therein). 
Also, at redshifts higher than six, quasars are found
that possess SMBHs with the mass 
higher than $10^9 ~M_\odot$ (Fan et al. 2001; Kurk et al., 2007). 
The formation history of these SMBHs is 
among the most significant 
unsolved issues in astrophysics (Volonteri 2012; Haiman 2013). 
Two recently discovered high-redshift quasars, 
ULASJ112010+641 with the mass of $M_\mathrm{BH} = 2 \times 10^9 ~M_\odot$ at redshift $z=7.085$  (Mortlock et al. 2011)
and SDSS J01001+2802 with $M_\mathrm{BH} = 1.2 \times 10^{10} ~M_\odot$ at $z=6.30$ (Wu et al. 2015) 
have raised a serious problem for the formation of SMBHs. 
Possible building blocks for such high-redshift SMBHs are the remnants of first stars. 
The initial mass function of first stars is thought to be more or less top-heavy 
(Abel et al. 2000; Nakamura \& Umemura 2001; Bromm et al. 2002; Yoshida et al. 2006; 
Greif et al. 2011; Susa et al. 2014; Hirano et al 2014).
First stars of several tens $M_\odot$ undergo supernovae, leaving black holes (BHs) 
of few tens $M_\odot$ (Heger \& Wooseley 2002). 
For SMBHs 
to grow from such first star remnants through mass accretion at  $z \gtrsim 6$, 
a super-Eddington accretion rate is requisite.  
If SMBHs grow continuously by mass accretion from BH remnants of $\sim 20 M_\odot$, the Eddington ratio ($\lambda$) 
is required to be $\lambda =1.4$ for ULASJ112010+641, or $\lambda =1.3$ for SDSS J01001+2802.
However, the continuous accretion is unlikely to be sustained due to feedbacks, and thus 
the average mass accretion rates should be lower than the Eddington rate 
 (Milosavljevic 2009; Alvarez et al 2009). On the other hand,
seed BHs may stem from supermasssive stars of $10^{4-6} ~M_\odot$
as a result of the direct collapse of primordial density fluctuations
 (Umemura, Loeb \& Turner 1993; Bromm \& Loeb 2003; Inayoshi \& Omukai 2012).  
 These BHs are thought  to be incorporated into a primordial galaxy of  $\sim 10^8-10^9 ~{M_\odot}$
(Greene 2012). 
If an SMBH grows via gas accretion from such a massive BH, the constraint on the accretion rate 
can be alleviated. 

Another possible pathway of SMBH formation is the merger of BHs. 
If the merger of multiple black holes precedes the growth via gas accretion,
the merged BH can be a seed of a supermassive black hole. 
So far, the merger of SMBHs in a galaxy has been argued extensively.
As for a binary of SMBHs, Begelman et al. (1980) pointed out that
the orbit of a binary SMBH 
cannot shrink below one parsec due to the loss cone depletion (the depletion of stars on orbits 
that intersect the binary SMBH), which is often called the final parsec problem (e.g. Merritt et al. 2004).  
It is argued that if the host galaxy provides an aspherical potential, a binary SMBH may overcome
the final parsec problem  
 (Khan, Just \& Merritt 2011, Khan et al. 2012; Khan et al. 2013). 
However, this solution of the problem is still under debate (Vasiliev, Antonini \& Merritt 2014). 
If there are more than two SMBHs in a galaxy, the dynamical relaxation of SMBHs is significantly 
controlled by the gravity of SMBHs themselves, especially by three-body interaction.  
When a third MBH intrudes into a binary SMBH, 
one SMBH 
carries away angular momentum from 
the remaining two SMBHs,
reducing the binary separation and eventually inducing the merger of the binary (Iwasawa, Funato \& Makino 2006).
In the case of  many SMBHs, the stellar dynamical friction allows a binary MBH to interact 
frequently with other SMBHs, and then the decay of the binary orbits leads to the merger
(Tanikawa \& Umemura 2011, 2014).

In a first-generation object formed at an early cosmic epoch,
the dynamical friction by stars is unlikely to work effectively,
since the initial mass function is top-heavy and most stars undergo supernovae. 
However, the dynamical friction by gas could work.
Recent radiation hydrodynamic simulations on the formation of first stars show 
that several or more stars form in a primordial gas cloud with the density of around 
$10^7 ~\mathrm{cm^{-3}}$ and the extension of $1000 ~\mathrm{AU}$,
where the gas fraction is $99\%$
(Greif et al. 2011; Umemura et al. 2012; Susa 2013; Susa et al. 2014) . 
In this circumstance, BH remnants of first stars are most likely subject 
to the dynamical friction by abundant gas. 
The gas dynamical friction has been considered as a mechanism that prompts 
the BH merger  (Ostriker 1999, Tanaka \& Haiman 2007).
Hitherto, the merger processes by the gas dynamical friction have been investigated
in the case of two massive BHs (e.g. Escala 2004, Escala 2005).
In this paper, we explore the merger of multiple BHs,
supposing a first-generation object
of $\sim 10^5-10^6 ~{M_\odot}$ or a gas-rich primordial galaxy of $\sim 10^8-10^9 ~{M_\odot}$.

The paper is organized as follows: 
In section 2, we describe the numerical method. 
In section 3, we show the numerical results. 
In section 4, we discuss the merger criterion through the gas friction. 
In section 5, we summarize the paper. 

\section{Method of Numerical Simulations}
Here, we present the framework of numerical simulations.

\subsection{Equation of motion}

The equations of motion for BHs are given by
\begin{equation}
  \frac{d^2{\bf r}_i}{dt^2} = \sum_{j}^{N_{\rm BH}} 
 \left\{ - Gm_{j}\frac{{\bf r}_{i}-{\bf r}_{j}}{|{\bf r}_{j}-
 {\bf r}_{j}|^3} + {\bf a}_{{\rm PN},ij} \right\} 
+ {{\bf a}^{\rm gas}_{\mathrm{DF},i}} 
+ {\bf a}_{\mathrm{pot},i},
\end{equation}
where ${\bf r}_i$ and ${\bf r}_j$ are respectively the positions of $i$-th BH and $j$-th BH, 
$N_{\rm BH}$ is the number of BHs, 
$G$ is the gravitational constant, $m_j$ is the mass  of $j$-th BH, 
${\bf a}_{{\rm PN},ij}$ is the general relativistic acceleration of $j$-th BH on $i$-th BH
in the Post-Newtonian prescription, 
${{\bf a}^{\rm gas}_{\mathrm{DF},i}}$ is the dynamical friction (DF) on $i$-th BH by gas,
and  ${\bf a}_{\mathrm{pot},i}$ is the acceleration on $i$-th BH by gravitational  potential of gas.

\subsection{Key parameters}
We initially set up multiple BHs of equal mass. In this paper,
we consider two cases of the BH mass: one is $30 ~M_\odot$ BHs as first star remnants, which are born
in a first-generation object of  $\sim 10^5-10^6 ~{M_\odot}$, and the other is 
$10^4 ~M_\odot$ BHs resulting from supermassive stars, which
are incorporated into a primordial galaxy of  $\sim 10^8-10^9 ~{M_\odot}$.
A key parameter in the present simulations is the BH density, $\rho_\mathrm{BH}$, at the initial epoch. 
Recent radiation hydrodynamic simulations on first star formation have shown that 
several or more stars are born in a disk of $\sim 1000 ~\mathrm{AU}$ (Greif et al. 2011; Umemura et al. 2012; Susa 2013; Susa et al 2014).
We change the typical extension of the BH distribution at the initial epoch, $r_\mathrm{typ}$, to settle the BH density. 
In the case of $30 ~M_\odot$ mass BH, we alter $r_\mathrm{typ}$ from $0.01 ~\mathrm{pc}$ to $1 ~\mathrm{pc}$.
In the case of $10^4 ~M_\odot$ mass BH, we set $r_\mathrm{typ}$ from $0.1 ~\mathrm{pc}$ 
to $10 ~\mathrm{pc}$. Then, the BH density is given by
\begin{equation}
	\rho_\mathrm{BH}=\frac{3}{4\pi r_\mathrm{typ}^3}\sum_{i} m_{i},
\end{equation}
where $m_i$ is the mass of $i$-th BH.
Another key parameter is the gas number density $n_\mathrm{gas}$. 
Simulations of first stars show that the density in a first object is $\sim 10^{7-8} ~\mathrm{cm^{-3}}$ 
in the star-formation epoch.
Here, we consider a wider range of the gas density from $10^4 ~\mathrm{cm^{-3}}$ to $10^{12} ~\mathrm{cm^{-3}}$
to elucidate the dependence of the BH merger on the gas density.

\subsection{Gas drag and potential}
We give the gas dynamical friction and the gas potential by analytic solutions.
We use the formula of the gas dynamical friction force given by Tanaka \& Haiman (2009) 
for the motion with ${\cal M}_i<{\cal M}_{eq}$ and  Ostriker (1999) 
for ${\cal M}_i>{\cal M}_{eq}$, where ${\cal M}_i$ is the Mach number of $i$-th BH and ${\cal M}_{eq}$ is 
the Mach number where these two formulas give equal acceleration.
Here, we adopt ${\cal M}_{eq}=1.5$ as Tanaka \& Haiman (2009). 
Then, the acceleration of the gas dynamical friction (${{\bf a}^{\rm gas}_{\mathrm{DF},i}}$) is given by
\begin{equation}
	{{\bf a}^{\rm gas}_{\mathrm{DF},i}} = - 4{\pi} {G}^2 m_i m_\mathrm{H}{n_\mathrm{gas}(r)}\frac{{\bf v}_i}{{v_i}^3} {\times} f({\cal M}_i) 
\end{equation}
\begin{equation}
f({\cal M}_i) =
\left \{
\begin{array}{ll}
	0.5 \mathrm{ln} \left( \frac{v_it}{{r}_\mathrm{min}} \right) \left[ \mathrm{erf} \left(\frac{{\cal M}_i}{\sqrt{2}} \right) - {\sqrt{\frac{2}{\pi}}} {\cal M}_i \mathrm{exp}( - \frac{{\cal M}_i ^2}{2}) \right]  ,\\\\
( 0 {\leq} {\cal M}_i {\leq} 0.8 ) \\\\
1.5 \mathrm{ln} \left( \frac{v_it}{{r}_\mathrm{min}} \right) \left[ \mathrm{erf} \left( \frac{{\cal M}_i}{\sqrt{2}} \right) - {\sqrt{\frac{2}{\pi}}} {{\cal M}_i} \mathrm{exp}( - \frac{{\cal M}_i ^2}{2}) \right]   ,\\\\
( 0.8 {\leq} {\cal M}_i {\leq} {\cal M}_{eq} ) \\\\
	\frac{1}{2} \mathrm{ln} \left( 1 - \frac{1}{{\cal M}_i^2} \right) + \mathrm{ln} \left( \frac{v_it}{{r}_\mathrm{min}} \right)  ,\\\\
({\cal M}_{eq} {\leq} {\cal M}_i )
\end{array}
\right. 
\end{equation}
where $m_\mathrm{H}$ is the mass of the hydrogen atom, $n_\mathrm{gas}$ is the number density of gas, 
$v_i$ is the velocity of $i$-th BH, and $t$ is the elapsed time. 
The $r_\mathrm{min}$ is the minimum scale of the dynamical friction on a BH, and we give $r_\mathrm{min}$ 
as $Gm_i/v_i^2$. Here, $v_it$ means the effective scale of gas medium, and we set an upper 
limit of $v_it$ to $0.1 ~\mathrm{pc}$. 
When $v_it < r_\mathrm{min}$, we assume $f({\cal M}_i)=0$.

In this paper, we postulate uniform background gas 
to purely extract the dependence on the gas density,
and give its density as a parameter. 
Then, the gravitational acceleration by gas
(${\bf a}_{{\rm pot},i}$) and its time derivative (${\bf \dot{a}}_{{\rm pot},i}$) are given as 
\begin{equation}
	{\bf a}_{\mathrm{pot},i}=-\frac{4}{3}\pi G m_\mathrm{H}n_\mathrm{gas} {\bf r}_i,
\end{equation}
\begin{equation}
	{\bf \dot{a}}_{\mathrm{pot},i}=-\frac{4}{3}\pi G m_\mathrm{H}n_\mathrm{gas} {\bf v}_i.
\end{equation}

\subsection{Gas temperature}
The heating rate due to the gas dynamical friction is estimated as follows  (Kim et al. 2005) :
\begin{eqnarray}
	\Lambda_\mathrm{DF}&=&7.3\times 10^{-25} ~\mathrm{erg~cm^{-3}~s^{-1}} 
	\left( \frac{n_\mathrm{gas}}{10^4~\mathrm{cm}^{-3}} \right) \nonumber\\
	&\times&\left(\frac{M_\mathrm{BH}}{30~M_\odot} \right)^2 
	\left(\frac{<f({\cal M})/\cal M>}{2} \right) \nonumber\\
	&\times&\left( \frac{T}{1000~ K} \right)^{-1/2}
	\left( \frac{n_\mathrm{BH}}{10~{\mathrm{pc}}^{-3}} \right) 
\end{eqnarray}
where $n_\mathrm{BH}$ is the number density of BHs, and the angular brackets denote 
the average over the Maxwellian distribution,
\begin{equation}
	f(v)=\frac{4\pi N_{\rm BH}}{(2\pi \sigma_r^2)^{3/2}}v^2e^{-v^2/(2\sigma_r^2)}, 
\end{equation}
where $\sigma_r$ is the one-dimensional velocity dispersion. 

In a first object, the cooling is dominated by hydrogen molecules (H$_2$) at $T \approx 10^3~\mathrm{K}$ (e.g. Omukai 2000).
In a primordial galaxy, the temperature goes down by the cooling of neutral hydrogen (HI) around $T \simeq 10^4~\mathrm{K}$,
and further reduces by the H$_2$ cooling down to $T \approx 10^3~\mathrm{K}$, if the gaseous metallicity is lower than
a percent of solar abundance (Susa \& Umemura 2004). 
The H$_2$ cooling rate through $\mathrm{H-H_2}$ collision (Hollenbach \& McKee 1989) 
and the HI cooling rate (Thoul \& Weinberg 1995) are respectively given by
\begin{eqnarray}
	\Lambda_\mathrm{H2}(10^3~K)&=&2.8\times 10^{-21} ~\mathrm{erg~cm^{-3}~s^{-1}} \nonumber\\
	&\times&\left( \frac{n_\mathrm{gas}}{10^4~\mathrm{cm}^{-3}} \right)
	\left(\frac{f_\mathrm{H_2}}{3\times10^{-4}} \right)
\end{eqnarray}
\begin{eqnarray}
	\Lambda_\mathrm{HI}(10^4~K)&\sim&1.0\times 10^{-14} ~\mathrm{erg~cm^{-3}~s^{-1}} \nonumber\\
	&\times&\left( \frac{n_\mathrm{gas}}{10^4~\mathrm{cm}^{-3}} \right)^2
	f_\mathrm{HI}^2
\end{eqnarray}
where 
$f_\mathrm{H_2}$ is the fraction of $\mathrm{H_2}$ molecules
and
$f_\mathrm{HI}$ is the fraction of neutral hydrogen atoms. 
Here, $f_\mathrm{H_2}~\gtrsim~3\times10^{-4}$ in the range of $n_\mathrm{gas}\gtrsim 10^4~\mathrm{cm}^{-3}$ 
(Palla, Salpeter \& Stahler 1983).  



We find that, in both cases of BH mass, the heating rate is lower than the cooling rate
over the range of parameters in which the simulations are performed. 
Therefore, the gas temperature is expected to settle at $T \approx 10^3~\mathrm{K}$. 
In this paper, we assume the gas temperature to be $1000 ~\mathrm{K}$.
Then, the sound speed is given as $C_\mathrm{s}=3.709 ~\mathrm{[km/s]}$.

\subsection{Relativistic effects}
We incorporate the general relativistic effects according to the Post Newtonian prescription up to 2.5PN term (Kupi 2006). 
The relativistic effects on $i$-th BH by $j$-th BH are given by 1PN (${\bf a}_\mathrm{1PN,ij}$) and 2PN  (${\bf a}_\mathrm{2PN,ij}$) term
corresponding to the pericentre shift, and 2.5PN term (${\bf a}_\mathrm{2.5PN,ij}$) corresponding to
the gravitational wave (GW) radiation. They are expressed by  
\begin{eqnarray}
{\bf a}_\mathrm{1PN,ij}
& = & \frac{G{m}_{j}}{r_{ij}^2} [ {\bf n} [ - {{\bf v}_{i}}^{2} - 2 {{\bf v}_{j}}^{2} + 4 {\bf v}_{i} {\bf v}_{j} + \frac{3}{2} ({\bf n} {\bf v}_{j})^{2}\nonumber \\
& &+ 5 \left(\frac{G{m}_{i}}{r_{ij}}\right) + 4 \left(\frac{G{m}_{j}}{r_{ij}}\right) ] \nonumber \\
& & + ({\bf v}_{i} - {\bf v}_{j}) (4 {\bf n} {\bf v}_{i} - 3 {\bf n} {\bf v}_{j} ) ] 
\end{eqnarray}
\begin{eqnarray}
 {\bf a}_\mathrm{2PN,ij}
& = & \frac{G{m}_{j}}{r_{ij}^2} [ {\bf n} [ - 2{{\bf v}_{j}}^4 + 4 {{\bf v}_{j}}^2({\bf v}_{i} {\bf v}_{j}) - ({\bf v}_{i} {\bf v}_{j})^2 \nonumber \\
& & + \frac{3}{2} {{\bf v}_{i}}^2 ({\bf n} {\bf v}_{j})^2 + \frac{9}{2} {{\bf v}_{j}}^2 ({\bf n} {\bf v}_{j})^2 - 6({\bf v}_{i} {\bf v}_{j})({\bf n} {\bf v}_{j} )^2\nonumber \\
& & - \frac{15}{8} ({\bf n} {\bf v}_{j})^4 + \left(\frac{G {m}_{i}}{r_{ij}}\right) [ - \frac{15}{4}{{\bf v}_{i}}^2 + \frac{5}{4}{{\bf v}_{j}}^2\nonumber \\ 
& & - \frac{5}{2} {\bf v}_{i} {\bf v}_{j} + \frac{39}{2}({\bf n} {\bf v}_{i})^2 -39({\bf n} {\bf v}_{i})({\bf n} {\bf v}_{j}) + \frac{17}{2} ({\bf n} {\bf v}_{i})^2 ] \nonumber \\
& & + \left(\frac{G {m}_{j}}{r_{ij}}\right) [ 4{{\bf v}_{j}}^2 + 8 {\bf v}_{i} {\bf v}_{j} + 2({\bf n} {\bf v}_{i})^2\nonumber \\
& & - 4({\bf n} {\bf v}_{i})({\bf n} {\bf v}_{j}) - 6({\bf n} {\bf v}_{i})^2 ] + ({\bf v}_{i} - {\bf v}_{j}) [ {{\bf v}_{i}}^2 ({\bf n} {\bf v}_{j})\nonumber \\
& & + 4 {{\bf v}_{j}}^2 ({\bf n} {\bf v}_{i}) -5 {{\bf v}_{j}}^2({\bf n} {\bf v}_{2}) - 4({\bf v}_{i} {\bf v}_{j})({\bf n} {\bf v}_{i})\nonumber \\ 
& & +4({\bf v}_{i} {\bf v}_{j})({\bf n} {\bf v}_{j}) - 6 ({\bf n} {\bf v}_{i})({\bf n} {\bf v}_{j})^2 + \frac{9}{2}({\bf n} {\bf v}_{j})^3\nonumber \\
& & + \left(\frac{G{m}_{i}}{r_{ij}}\right) \left(- \frac{63}{4} {\bf n} {\bf v}_{i} + \frac{55}{4} {\bf n} {\bf v}_{j}\right)\nonumber \\
& & + \left(\frac{G{m}_{j}}{r_{ij}}\right)(- 2 {\bf n} {\bf v}_{i} - 2 {\bf n} {\bf v}_{j}) ] ] \nonumber \\
& & + \frac{G^3{m}_{j}}{r_{ij}^4} {\bf n} [ - \frac{57}{4} {{m}_{i}}^2 - 9{{m}_{j}}^2 - \frac{69}{2} {m}_{i}{m}_{j} ] 
\end{eqnarray}
\begin{eqnarray}
{\bf{a}}_\mathrm{2.5PN,ij}
& = & \frac{4}{5} \frac{G^2{m}_{i}{m}_{j}}{r_{ij}^3} [ ({{\bf v}}_{i} - {{\bf v}}_{j} )[ - ({{\bf v}}_{i} - {{\bf v}}_{j} )^2\nonumber \\
& & + 2 \left(\frac{G{m}_{i}}{r_{ij}} \right) - 8 \left(\frac{G{m}_{j}}{r_{ij}} \right) ]+ {\bf n} ( {\bf n} {{\bf v}}_{i} - {\bf n} {{\bf v}}_{2} ) \nonumber \\
& & [ 3({{\bf v}}_{i} - {{\bf v}}_{j} )^2 - 6 \left(\frac{G{m}_{i}}{r_{ij}} \right) + \frac{52}{3} \left( \frac{G{m}_{j}}{r_{ij}} \right) ] ] 
\end{eqnarray}
where
\begin{equation}
	{\bf n}=\frac{{\bf r}_{ij}}{r_{ij}},
\end{equation}
\begin{equation}
	{\bf r}_{ij}=({\bf r}_i-{\bf r}_j).
\end{equation}

\subsection{Merger condition}

We assume that two MBHs merge, when their separation is
less than 100 times the sum of their Schwarzschild radii:
\begin{equation}
  \left|\bm{r}_{i} - \bm{r}_{j} \right| < 100
  \left( r_{{\rm sch},i} + r_{{\rm sch},j} \right),
\end{equation}
where $r_{{\rm sch},i}$ is the Schwarzschild radius of $i$-th BH
given by $2Gm_{i}/c^2$ with the speed of light $c$.
At the final stage of merger, the binding energy of a binary BH is transformed to the energy of GW, 
retaining the mass of each BH. Hence, the BH mass after the merger is just the sum of two MBHs.

\subsection{Numerical scheme}
We integrate the equations of motion using the fourth-order Hermite scheme 
with the shared time step (Makino \& Aarseth 1992).
To use the Hermite scheme, 
we calculate the time derivative of the acceleration by the Newtonian gravity, the gas gravitational potential,
and the relativistic force. 
On the other hand, we treat the dynamical friction of gas to quadratic order, since
the formula of the dynamical friction is not so accurate as that of gravity
and also the fourth-order scheme is especially requisite during the gravitational three-body interaction 
between a close BH binary and an intruding BH. 
Therefore, the time derivative of the gas dynamical friction is not included.
Then, the shared time step in the Hermite scheme is given by
\begin{equation}
	\Delta t=\mathrm{min}_{i}\sqrt{\eta \frac{|{\bf a}_{\mathrm{Her},i}||{\bf a}_{\mathrm{Her},i}^{(2)}|+|{\bf a}_{\mathrm{Her},i}^{(1)}|^2}{|{\bf a}_{\mathrm{Her},i}^{(1)}||{\bf a}_{\mathrm{Her},i}^{(3)}|+|{\bf a}_{\mathrm{Her},i}^{(2)}|^2}},
\end{equation}
where ${\bf a}_{\mathrm{Her},i}$ is the acceleration on $i$-th BH that is treated to the fourth-order, and given by
\begin{equation}
	{\bf a}_{\mathrm{Her},i} = \sum_{j}^{N_{\rm BH}} 
 \left\{ - Gm_{j}\frac{{\bf r}_{i}-{\bf r}_{j}}{|{\bf r}_{j}-
 {\bf r}_{j}|^3} + {\bf a}_{{\rm PN},ij} \right\} 
+ {\bf a}_{\mathrm{pot},i}.
\end{equation}
The ${\bf a}_{\mathrm{Her},i}^{(k)}$ is the $k$-th derivative of ${\bf a}_{\mathrm{Her},i}$, 
and $\eta$ is the accuracy parameter. We assume $\eta =0.003$ in the present simulations. 

To avoid cancellation of significant digits when such tiny scales as $100~r_{{\rm sch}}$ are resolved, 
we calculate the BHs evolution in the coordinate 
where the origin is always set to the center of mass for the closest pair of BHs. 
This prescription is based on the simulations on the merger of multiple BHs 
(Tanikawa \& Umemura 2011), 
and allows us to pursue accurately the orbit of the BHs until the merger condition is satisfied.

\subsection{Setup of simulations}

We set up ten BHs as a fiducial case, and
besides investigate the cases of two or three BHs to scrutinize the key physics of merger.
The initial positions of BHs are given randomly in the $x-y$ plane within  
$r_\mathrm{typ}$. Also, we give the velocity to each BH as the sum of a circular component and a random 
component. The circular velocity is given to balance 
against the gravity in the $x-y$ plane. The random velocity is given according to a Gaussian distribution 
with the same dispersion as the circular velocity. The random velocity is given in the $xyz$ space.
In the case of two BHs, we give only circular velocity without initial eccentricity. 
This condition is desired to discriminate the BH merger purely by the gas friction.

We calculate each run until $100~\mathrm{Myr}$, since the background  environments 
of the host objects are likely to change in $100~\mathrm{Myr}$. 
Also, we terminate the simulation, if all BHs merge into one BH.

\begin{figure}
\includegraphics[width=80mm]{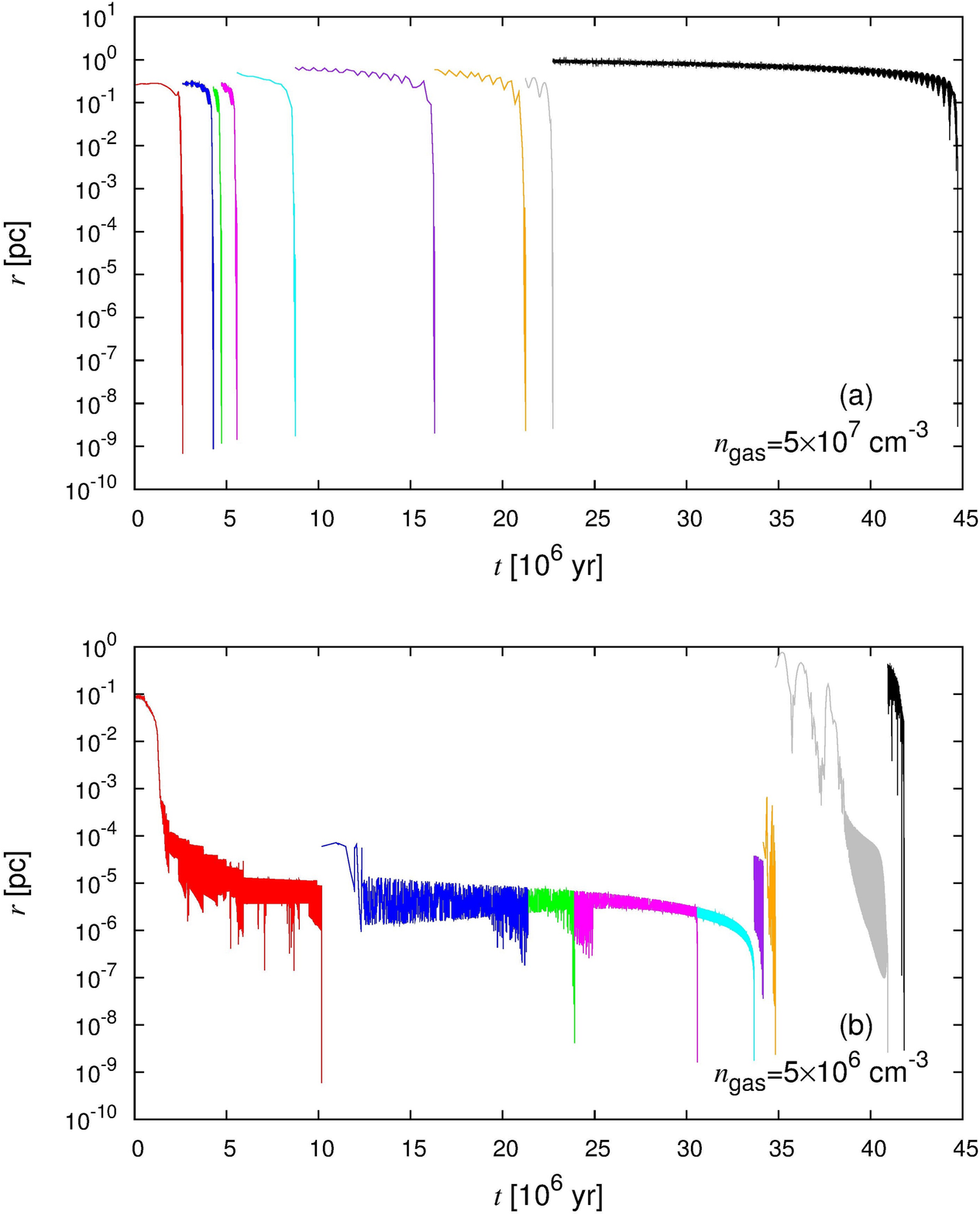}
\vspace{5mm}
\caption{The separation of the closest pair within all BHs as a function of time. 
The initial BH mass is $M_\mathrm{BH}=30 ~M_\odot$, the initial typical extension of the BH distribution is 
$r_\mathrm{typ}=1.0 ~\mathrm{pc}$, and the gas density is $n_\mathrm{gas}=5\times10^7$ (top) 
or $5\times10^6$ (bottom) $\mathrm{cm}^{-3}$.}
\label{30_1}
\end{figure}

\begin{figure}
	\begin{center}
\includegraphics[width=80mm]{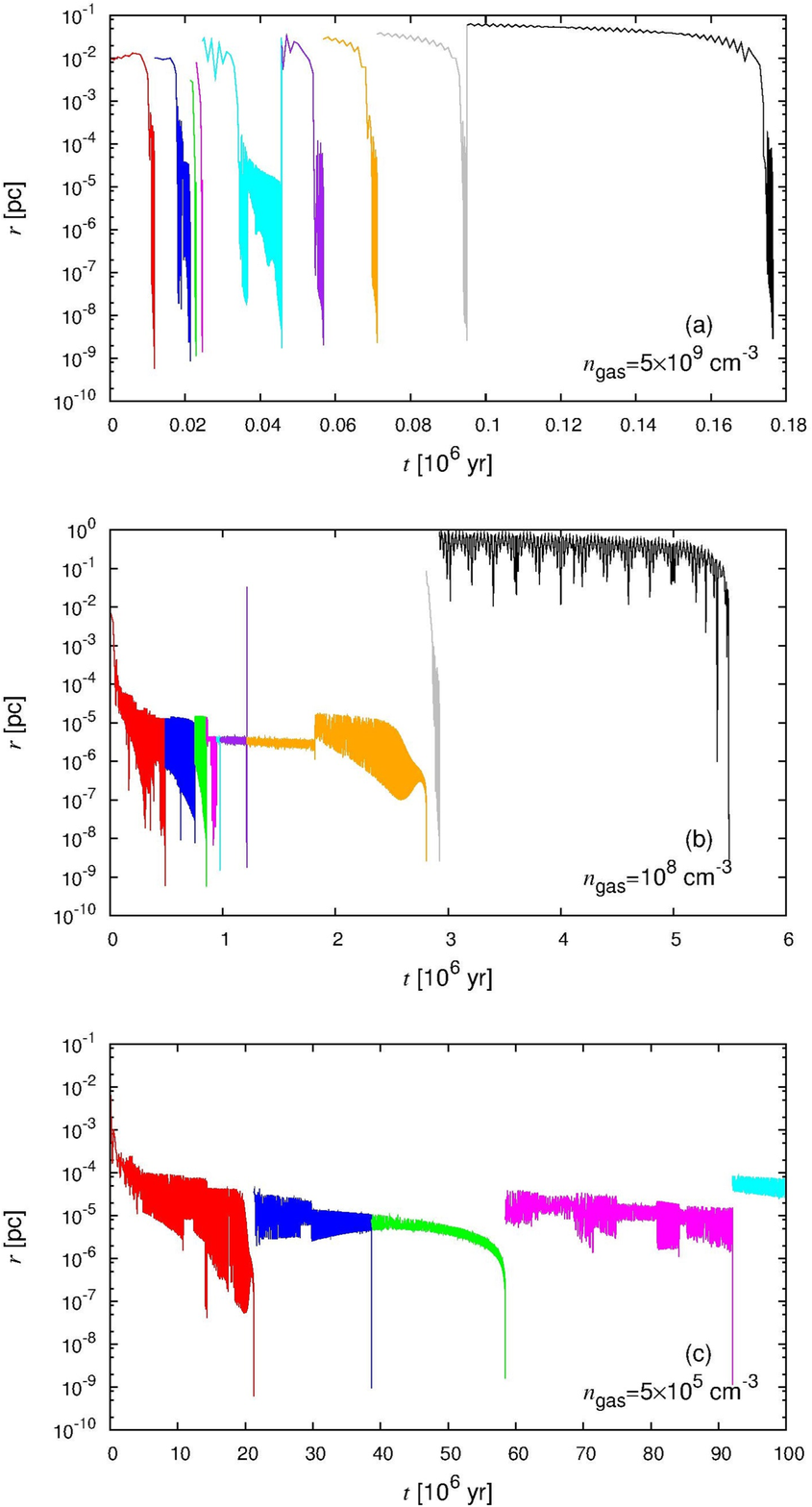}
\vspace{5mm}
\caption{Same as Figure \ref{30_1}, but for $r_\mathrm{typ}=0.1\mathrm{pc}$, and
$n_\mathrm{gas}=5\times10^9$ (top), $10^8$ (middle), or $5\times10^5$ (bottom) $\mathrm{cm}^{-3}$.}
\label{30_01}
\end{center}
\end{figure}

\begin{figure}
\begin{center}
\includegraphics[width=80mm]{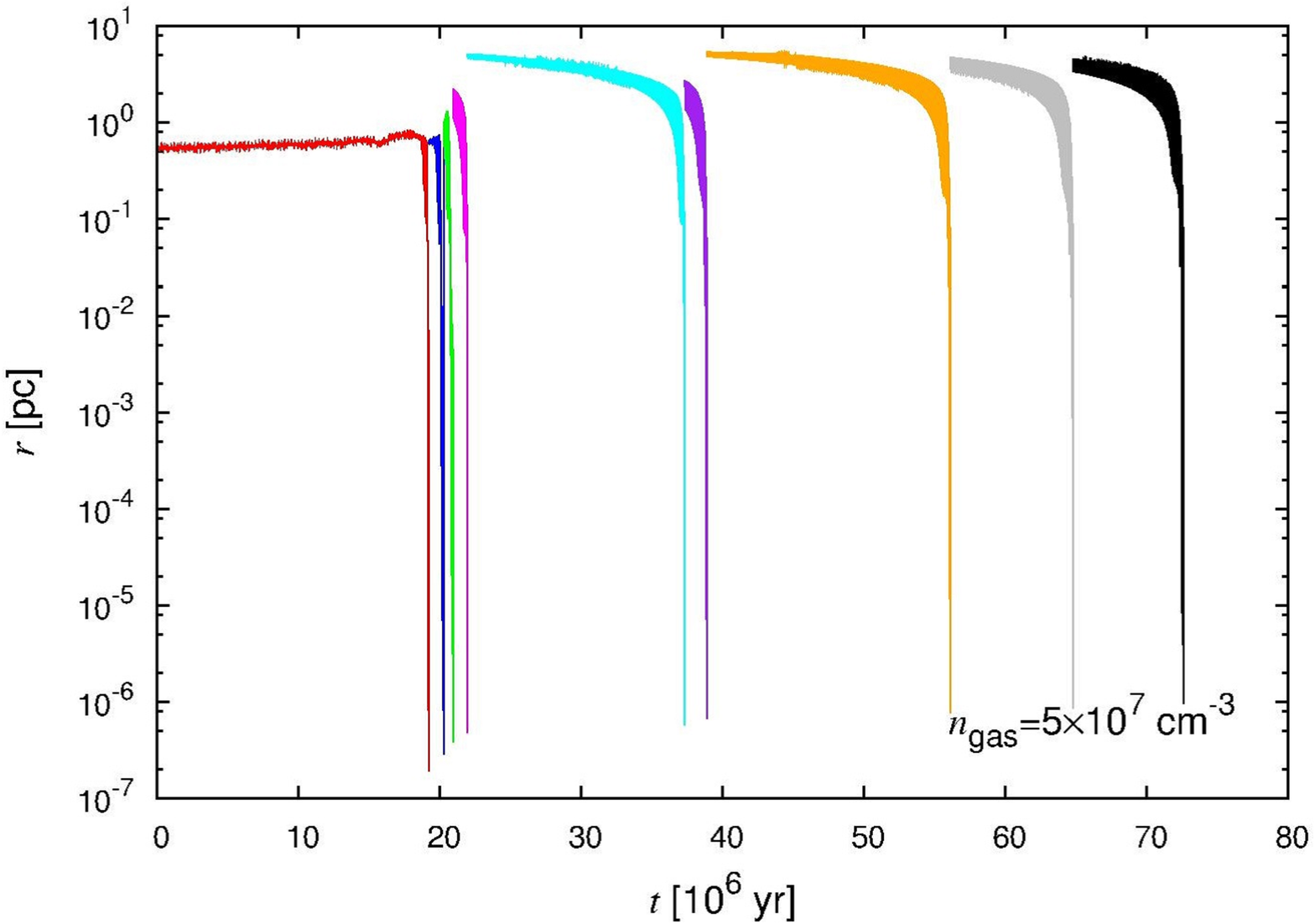}
\vspace{5mm}
\caption{Same as Figure \ref{30_1}, but for $M_\mathrm{BH}=10^4 ~M_\odot,r_\mathrm{typ}=10.0 ~\mathrm{pc}$, 
and $n_\mathrm{gas}=5\times10^7 ~\mathrm{cm}^{-3}$.}
\label{104_10}
\end{center}
\end{figure}

\begin{figure}
	\begin{center}
\includegraphics[width=80mm]{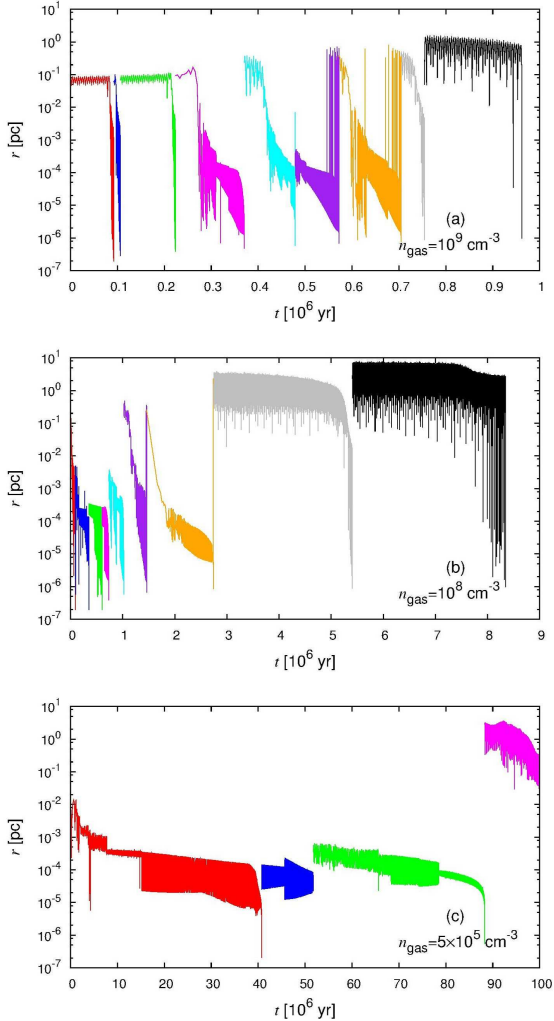}
\vspace{5mm}
\caption{Same as Figure \ref{30_1}, but for $M_\mathrm{BH}=10^4 ~M_\odot, r_\mathrm{typ}=1.0~\mathrm{pc}$, 
and $n_\mathrm{gas}=10^9$ (top), $10^8$ (middle), or $5\times10^5$ (bottom) $\mathrm{cm}^{-3}$.}
\label{104_1}
\end{center}
\end{figure}

\begin{figure}
	\begin{center}
\includegraphics[width=80mm]{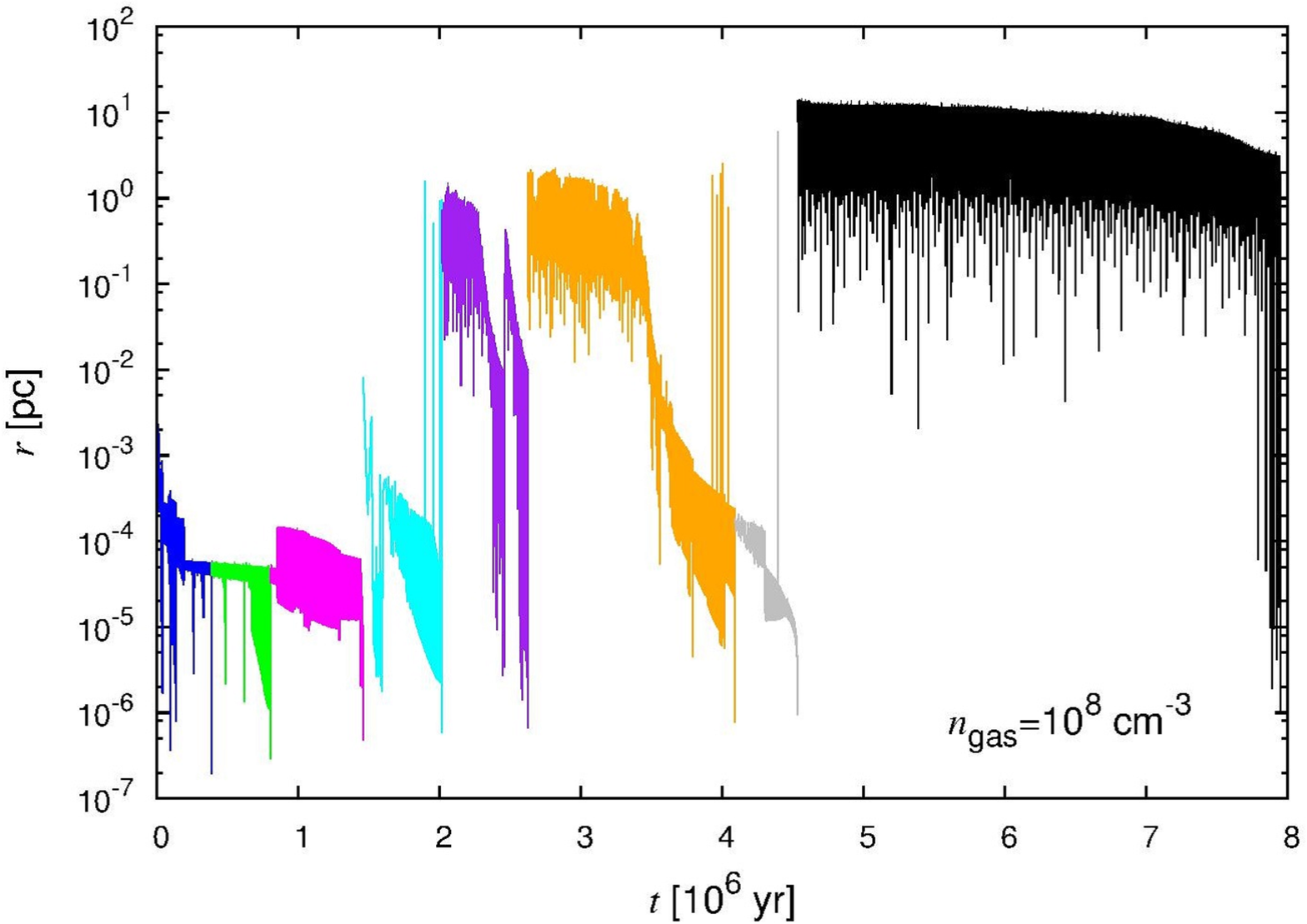}
\vspace{5mm}
\caption{Same as Figure \ref{30_1}, but for $M_\mathrm{BH}=10^4 ~M_\odot,r_\mathrm{typ}=0.1 ~\mathrm{pc}$, 
and $n_\mathrm{gas}=10^8 ~\mathrm{cm}^{-3}$.}
\label{104_01}
\end{center}
\end{figure}

\begin{table*}
	\begin{minipage}{120mm}
		\caption{The results with $M_\mathrm{BH}=30~M_\odot$ and $N_{\rm BH}=10$.}
		\label{res_t1}
\begin{tabular}{|c|c|c|c|c|c|c|c|c|c|c|c|c|c|c|}
\hline \hline
$r_{\rm typ} [\mathrm{pc}]$& \multicolumn{2}{c|}{1.0} & \multicolumn{2}{c|}{0.464}& \multicolumn{2}{c|}{0.215} & \multicolumn{2}{c|}{0.1}& \multicolumn{2}{c|}{0.0464} & \multicolumn{2}{c|}{0.0215}& \multicolumn{2}{c|}{0.01} \\
\hline
$\rho_{\rm BH} [M_\odot \mathrm{pc^{-3}}]$& \multicolumn{2}{c|}{$7.2\times 10^{1}$} & \multicolumn{2}{c|}{$7.2\times 10^{2}$}& \multicolumn{2}{c|}{$7.2\times 10^{3}$} & \multicolumn{2}{c|}{$7.2\times 10^{4}$}& \multicolumn{2}{c|}{$7.2\times 10^{5}$} & \multicolumn{2}{c|}{$7.2\times 10^{6}$}& \multicolumn{2}{c|}{$7.2\times 10^{7}$} \\
\hline \hline
&$N_\mathrm{m}$&type&$N_\mathrm{m}$&type&$N_\mathrm{m}$&type&$N_\mathrm{m}$&type&$N_\mathrm{m}$&type&$N_\mathrm{m}$&type&$N_\mathrm{m}$&type\\
\cline{2-15} 
$n_\mathrm{gas} [\mathrm{cm}^{-3}]$  & \multicolumn{2}{c|}{$t_\mathrm{fin} [\mathrm{yr}]$} & \multicolumn{2}{c|}{$t_\mathrm{fin} [\mathrm{yr}]$}& \multicolumn{2}{c|}{$t_\mathrm{fin} [\mathrm{yr}]$} & \multicolumn{2}{c|}{$t_\mathrm{fin} [\mathrm{yr}]$}& \multicolumn{2}{c|}{$t_\mathrm{fin} [\mathrm{yr}]$} & \multicolumn{2}{c|}{$t_\mathrm{fin} [\mathrm{yr}]$}& \multicolumn{2}{c|}{$t_\mathrm{fin} [\mathrm{yr}]$} \\
\hline \hline
&0&-&5&A&{\bf 9}&A&{\bf 9}&A&{\bf 9}&A&{\bf 9}&A&{\bf 9}&B\\
\cline{2-15}
$10^{12}$ & \multicolumn{2}{c|}{$1.0\times10^8$} & \multicolumn{2}{c|}{$1.0\times10^8$}& \multicolumn{2}{c|}{$3.6\times10^7$} & \multicolumn{2}{c|}{$4.4\times10^6$}& \multicolumn{2}{c|}{$1.7\times10^6$} & \multicolumn{2}{c|}{$7.0\times10^5$}& \multicolumn{2}{c|}{$4.3\times10^5$} \\
\hline
&2&A&8&A&{\bf 9}&A&{\bf 9}&A&{\bf 9}&A&{\bf 9}&A&{\bf 9}&B\\
\cline{2-15}
$10^{11}$ & \multicolumn{2}{c|}{$1.0\times10^8$} & \multicolumn{2}{c|}{$1.0\times10^8$}& \multicolumn{2}{c|}{$1.3\times10^7$} & \multicolumn{2}{c|}{$1.8\times10^5$}& \multicolumn{2}{c|}{$1.8\times10^5$} & \multicolumn{2}{c|}{$2.3\times10^4$}& \multicolumn{2}{c|}{$2.8\times10^4$} \\
\hline
&5&A&{\bf 9}&A&{\bf 9}&A&{\bf 9}&A&{\bf 9}&A&{\bf 9}&B&{\bf 9}&B\\
\cline{2-15}
$10^{10}$ & \multicolumn{2}{c|}{$1.0\times10^8$} & \multicolumn{2}{c|}{$4.5\times10^7$}& \multicolumn{2}{c|}{$5.1\times10^6$} & \multicolumn{2}{c|}{$5.9\times10^5$}& \multicolumn{2}{c|}{$7.4\times10^4$} & \multicolumn{2}{c|}{$6.3\times10^4$}& \multicolumn{2}{c|}{$1.7\times10^5$} \\
\hline
&5&A&{\bf 9}&A&{\bf 9}&A&{\bf 9}&A&{\bf 9}&B&{\bf 9}&B&{\bf 9}&B\\
\cline{2-15}
$5\times10^9$ & \multicolumn{2}{c|}{$1.0\times10^8$} & \multicolumn{2}{c|}{$2.9\times10^7$}& \multicolumn{2}{c|}{$3.1\times10^6$} & \multicolumn{2}{c|}{$4.5\times10^5$}& \multicolumn{2}{c|}{$1.6\times10^5$} & \multicolumn{2}{c|}{$3.4\times10^5$}& \multicolumn{2}{c|}{$7.7\times10^4$} \\
\hline
&8&A&{\bf 9}&A&{\bf 9}&A&{\bf 9}&B&{\bf 9}&B&{\bf 9}&B&{\bf 9}&B\\
\cline{2-15}
$10^9$ & \multicolumn{2}{c|}{$1.0\times10^8$} & \multicolumn{2}{c|}{$2.4\times10^7$}& \multicolumn{2}{c|}{$3.5\times10^6$} & \multicolumn{2}{c|}{$2.6\times10^5$}& \multicolumn{2}{c|}{$4.3\times10^5$} & \multicolumn{2}{c|}{$3.5\times10^5$}& \multicolumn{2}{c|}{$3.0\times10^5$} \\
\hline
&8&A&{\bf 9}&A&{\bf 9}&A&{\bf 9}&B&{\bf 9}&B&{\bf 9}&B&{\bf 9}&B\\
\cline{2-15}
$5\times10^8$ & \multicolumn{2}{c|}{$1.0\times10^8$} & \multicolumn{2}{c|}{$1.2\times10^7$}& \multicolumn{2}{c|}{$1.3\times10^6$} & \multicolumn{2}{c|}{$6.5\times10^5$}& \multicolumn{2}{c|}{$4.3\times10^5$} & \multicolumn{2}{c|}{$5.5\times10^5$}& \multicolumn{2}{c|}{$5.1\times10^5$} \\
\hline
&{\bf 9}&A&{\bf 9}&A&{\bf 9}&B&{\bf 9}&B&{\bf 9}&B&{\bf 9}&B&{\bf 9}&B\\
\cline{2-15}
$10^8$ & \multicolumn{2}{c|}{$3.0\times10^7$} & \multicolumn{2}{c|}{$5.1\times10^6$}& \multicolumn{2}{c|}{$4.0\times10^6$} & \multicolumn{2}{c|}{$5.5\times10^6$}& \multicolumn{2}{c|}{$3.6\times10^6$} & \multicolumn{2}{c|}{$4.2\times10^6$}& \multicolumn{2}{c|}{$1.2\times10^7$} \\
\hline
&{\bf 9}&A&{\bf 9}&B&{\bf 9}&B&{\bf 9}&B&{\bf 9}&B&{\bf 9}&B&{\bf 9}&B\\
\cline{2-15}
$5\times10^7$ & \multicolumn{2}{c|}{$4.5\times10^7$} & \multicolumn{2}{c|}{$3.7\times10^6$}& \multicolumn{2}{c|}{$2.2\times10^7$} & \multicolumn{2}{c|}{$3.2\times10^7$}& \multicolumn{2}{c|}{$1.3\times10^7$} & \multicolumn{2}{c|}{$4.7\times10^6$}& \multicolumn{2}{c|}{$3.6\times10^6$} \\
\hline
&{\bf 9}&B&{\bf 9}&B&{\bf 9}&B&{\bf 9}&B&{\bf 9}&B&{\bf 9}&B&{\bf 9}&B\\
\cline{2-15}
$10^7$ & \multicolumn{2}{c|}{$3.8\times10^7$} & \multicolumn{2}{c|}{$2.3\times10^7$}& \multicolumn{2}{c|}{$1.7\times10^7$} & \multicolumn{2}{c|}{$3.3\times10^7$}& \multicolumn{2}{c|}{$1.8\times10^7$} & \multicolumn{2}{c|}{$2.9\times10^7$}& \multicolumn{2}{c|}{$1.7\times10^7$} \\
\hline
&{\bf 9}&B&{\bf 9}&B&{\bf 9}&B&{\bf 9}&B&{\bf 9}&B&{\bf 9}&B&{\bf 9}&B\\
\cline{2-15}
$5\times10^6$ & \multicolumn{2}{c|}{$4.2\times10^7$} & \multicolumn{2}{c|}{$3.9\times10^7$}& \multicolumn{2}{c|}{$4.2\times10^7$} & \multicolumn{2}{c|}{$4.7\times10^7$}& \multicolumn{2}{c|}{$6.3\times10^7$} & \multicolumn{2}{c|}{$3.5\times10^7$}& \multicolumn{2}{c|}{$3.1\times10^7$} \\
\hline
&6&B&6&B&8&B&6&C&8&C&6&C&6&C\\
\cline{2-15}
$10^6$ & \multicolumn{2}{c|}{$1.0\times10^8$} & \multicolumn{2}{c|}{$1.0\times10^8$}& \multicolumn{2}{c|}{$1.0\times10^8$} & \multicolumn{2}{c|}{$1.0\times10^8$}& \multicolumn{2}{c|}{$1.0\times10^8$} & \multicolumn{2}{c|}{$1.0\times10^8$}& \multicolumn{2}{c|}{$1.0\times10^8$} \\
\hline
&2&C&6&C&6&C&4&C&5&C&3&C&4&C\\
\cline{2-15}
$5\times10^5$ & \multicolumn{2}{c|}{$1.0\times10^8$} & \multicolumn{2}{c|}{$1.0\times10^8$}& \multicolumn{2}{c|}{$1.0\times10^8$} & \multicolumn{2}{c|}{$1.0\times10^8$}& \multicolumn{2}{c|}{$1.0\times10^8$} & \multicolumn{2}{c|}{$1.0\times10^8$}& \multicolumn{2}{c|}{$1.0\times10^8$} \\
\hline
&0&-&0&-&1&C&0&-&1&C&2&C&0&-\\
\cline{2-15}
$10^5$ & \multicolumn{2}{c|}{$1.0\times10^8$} & \multicolumn{2}{c|}{$1.0\times10^8$}& \multicolumn{2}{c|}{$1.0\times10^8$} & \multicolumn{2}{c|}{$1.0\times10^8$}& \multicolumn{2}{c|}{$1.0\times10^8$} & \multicolumn{2}{c|}{$1.0\times10^8$}& \multicolumn{2}{c|}{$1.0\times10^8$} \\
\hline
&0&-&0&-&0&-&1&C&0&-&0&-&0&-\\
\cline{2-15}
$10^4$ & \multicolumn{2}{c|}{$1.0\times10^8$} & \multicolumn{2}{c|}{$1.0\times10^8$}& \multicolumn{2}{c|}{$1.0\times10^8$} & \multicolumn{2}{c|}{$1.0\times10^8$}& \multicolumn{2}{c|}{$1.0\times10^8$} & \multicolumn{2}{c|}{$1.0\times10^8$}& \multicolumn{2}{c|}{$1.0\times10^8$} \\
\hline
\end{tabular}
\end{minipage}
\end{table*}

\begin{table*}
	\begin{minipage}{120mm}
		\caption{The results with $M_\mathrm{BH}=10^4 ~M_\odot$ and $N_{\rm BH}=10$.}
			\label{res_t2}
\begin{tabular}{|c|c|c|c|c|c|c|c|c|c|c|c|c|c|c|}
\hline \hline
$r_{\rm typ} [\mathrm{pc}]$& \multicolumn{2}{c|}{10.0} & \multicolumn{2}{c|}{4.64}& \multicolumn{2}{c|}{2.15}& \multicolumn{2}{c|}{1.0} & \multicolumn{2}{c|}{0.464}& \multicolumn{2}{c|}{0.215} & \multicolumn{2}{c|}{0.1} \\
\hline
$\rho_{\rm BH} [M_\odot \mathrm{pc^{-3}}]$& \multicolumn{2}{c|}{$2.4\times 10^{1}$} & \multicolumn{2}{c|}{$2.4\times 10^{2}$}& \multicolumn{2}{c|}{$2.4\times 10^{3}$} & \multicolumn{2}{c|}{$2.4\times 10^{4}$}& \multicolumn{2}{c|}{$2.4\times 10^{5}$} & \multicolumn{2}{c|}{$2.4\times 10^{6}$}& \multicolumn{2}{c|}{$2.4\times 10^{7}$} \\
\hline \hline
&$N_\mathrm{m}$&type&$N_\mathrm{m}$&type&$N_\mathrm{m}$&type&$N_\mathrm{m}$&type&$N_\mathrm{m}$&type&$N_\mathrm{m}$&type&$N_\mathrm{m}$&type\\
\cline{2-15} 
$n_\mathrm{gas} [\mathrm{cm}^{-3}]$  & \multicolumn{2}{c|}{$t_\mathrm{fin} [\mathrm{yr}]$} & \multicolumn{2}{c|}{$t_\mathrm{fin} [\mathrm{yr}]$}& \multicolumn{2}{c|}{$t_\mathrm{fin} [\mathrm{yr}]$} & \multicolumn{2}{c|}{$t_\mathrm{fin} [\mathrm{yr}]$}& \multicolumn{2}{c|}{$t_\mathrm{fin} [\mathrm{yr}]$} & \multicolumn{2}{c|}{$t_\mathrm{fin} [\mathrm{yr}]$}& \multicolumn{2}{c|}{$t_\mathrm{fin} [\mathrm{yr}]$} \\
\hline \hline
&0&-&4&A&7&A&{\bf 9}&A&{\bf 9}&A&{\bf 9}&A&{\bf 9}&A\\
\cline{2-15}
$10^{12}$ & \multicolumn{2}{c|}{$1.0\times10^8$} & \multicolumn{2}{c|}{$\times10^{10}$}& \multicolumn{2}{c|}{$\times10^{10}$} & \multicolumn{2}{c|}{$1.3\times10^7$}& \multicolumn{2}{c|}{$1.5\times10^6$} & \multicolumn{2}{c|}{$1.7\times10^5$}& \multicolumn{2}{c|}{$2.2\times10^4$} \\
\hline
&0&-&5&A&{\bf 9}&A&{\bf 9}&A&{\bf 9}&A&{\bf 9}&A&{\bf 9}&B\\
\cline{2-15}
$10^{11}$ & \multicolumn{2}{c|}{$1.0\times10^8$} & \multicolumn{2}{c|}{$1.0\times10^8$}& \multicolumn{2}{c|}{$4.6\times10^7$} & \multicolumn{2}{c|}{$5.0\times10^6$}& \multicolumn{2}{c|}{$5.8\times10^5$} & \multicolumn{2}{c|}{$7.2\times10^4$}& \multicolumn{2}{c|}{$1.6\times10^4$} \\
\hline
&4&A&8&A&{\bf 9}&A&{\bf 9}&A&{\bf 9}&A&{\bf 9}&A&{\bf 9}&B\\
\cline{2-15}
$10^{10}$ & \multicolumn{2}{c|}{$1.0\times10^8$} & \multicolumn{2}{c|}{$1.0\times10^8$}& \multicolumn{2}{c|}{$1.7\times10^7$} & \multicolumn{2}{c|}{$2.0\times10^6$}& \multicolumn{2}{c|}{$2.5\times10^5$} & \multicolumn{2}{c|}{$9.3\times10^4$}& \multicolumn{2}{c|}{$4.1\times10^4$} \\
\hline
&5&A&{\bf 9}&A&{\bf 9}&A&{\bf 9}&A&{\bf 9}&B&{\bf 9}&B&{\bf 9}&B\\
\cline{2-15}
$10^9$ & \multicolumn{2}{c|}{$1.0\times10^8$} & \multicolumn{2}{c|}{$5.8\times10^7$}& \multicolumn{2}{c|}{$7.0\times10^6$} & \multicolumn{2}{c|}{$9.6\times10^5$}& \multicolumn{2}{c|}{$7.6\times10^5$} & \multicolumn{2}{c|}{$9.0\times10^5$}& \multicolumn{2}{c|}{$2.4\times10^6$} \\
\hline
&5&A&{\bf 9}&A&{\bf 9}&A&{\bf 9}&B&{\bf 9}&B&{\bf 9}&B&{\bf 9}&B\\
\cline{2-15}
$5\times10^8$ & \multicolumn{2}{c|}{$1.0\times10^8$} & \multicolumn{2}{c|}{$5.0\times10^7$}& \multicolumn{2}{c|}{$1.0\times10^7$} & \multicolumn{2}{c|}{$1.5\times10^6$}& \multicolumn{2}{c|}{$9.5\times10^5$} & \multicolumn{2}{c|}{$7.6\times10^5$}& \multicolumn{2}{c|}{$6.7\times10^5$} \\
\hline
&7&A&{\bf 9}&A&{\bf 9}&B&{\bf 9}&B&{\bf 9}&B&{\bf 9}&B&{\bf 9}&B\\
\cline{2-15}
$10^8$ & \multicolumn{2}{c|}{$1.0\times10^8$} & \multicolumn{2}{c|}{$2.5\times10^7$}& \multicolumn{2}{c|}{$3.7\times10^6$} & \multicolumn{2}{c|}{$8.3\times10^6$}& \multicolumn{2}{c|}{$4.9\times10^6$} & \multicolumn{2}{c|}{$5.6\times10^6$}& \multicolumn{2}{c|}{$7.9\times10^6$} \\
\hline
&{\bf 9}&A&{\bf 9}&A&{\bf 9}&B&{\bf 9}&B&{\bf 9}&B&{\bf 9}&B&{\bf 9}&B\\
\cline{2-15}
$5\times10^7$ & \multicolumn{2}{c|}{$7.3\times10^7$} & \multicolumn{2}{c|}{$2.0\times10^7$}& \multicolumn{2}{c|}{$1.3\times10^7$} & \multicolumn{2}{c|}{$7.1\times10^6$}& \multicolumn{2}{c|}{$1.2\times10^7$} & \multicolumn{2}{c|}{$3.5\times10^6$}& \multicolumn{2}{c|}{$9.4\times10^6$} \\
\hline
&{\bf 9}&B&{\bf 9}&B&{\bf 9}&B&{\bf 9}&B&{\bf 9}&B&{\bf 9}&B&{\bf 9}&B\\
\cline{2-15}
$10^7$ & \multicolumn{2}{c|}{$7.7\times10^7$} & \multicolumn{2}{c|}{$3.5\times10^7$}& \multicolumn{2}{c|}{$9.3\times10^7$} & \multicolumn{2}{c|}{$6.0\times10^7$}& \multicolumn{2}{c|}{$3.3\times10^7$} & \multicolumn{2}{c|}{$3.6\times10^7$}& \multicolumn{2}{c|}{$4.1\times10^7$} \\
\hline
&8&B&{\bf 9}&B&{\bf 9}&B&{\bf 9}&B&{\bf 9}&B&{\bf 9}&B&{\bf 9}&B\\
\cline{2-15}
$5\times10^6$ & \multicolumn{2}{c|}{$1.0\times10^8$} & \multicolumn{2}{c|}{$8.3\times10^7$}& \multicolumn{2}{c|}{$3.9\times10^7$} & \multicolumn{2}{c|}{$8.5\times10^7$}& \multicolumn{2}{c|}{$4.6\times10^7$} & \multicolumn{2}{c|}{$4.8\times10^7$}& \multicolumn{2}{c|}{$7.9\times10^7$} \\
\hline
&5&B&7&B&4&B&{\bf 9}&B&5&C&5&C&6&C\\
\cline{2-15}
$10^6$ & \multicolumn{2}{c|}{$1.0\times10^8$} & \multicolumn{2}{c|}{$1.0\times10^8$}& \multicolumn{2}{c|}{$1.0\times10^8$} & \multicolumn{2}{c|}{$6.5\times10^7$}& \multicolumn{2}{c|}{$1.0\times10^8$} & \multicolumn{2}{c|}{$1.0\times10^8$}& \multicolumn{2}{c|}{$1.0\times10^8$} \\
\hline
&3&B&3&C&3&C&3&C&6&C&4&C&4&C\\
\cline{2-15}
$5\times10^5$ & \multicolumn{2}{c|}{$1.0\times10^8$} & \multicolumn{2}{c|}{$1.0\times10^8$}& \multicolumn{2}{c|}{$1.0\times10^8$} & \multicolumn{2}{c|}{$1.0\times10^8$}& \multicolumn{2}{c|}{$1.0\times10^8$} & \multicolumn{2}{c|}{$1.0\times10^8$}& \multicolumn{2}{c|}{$1.0\times10^8$} \\
\hline
&0&-&0&-&0&-&0&-&0&-&1&C&0&-\\
\cline{2-15}
$10^5$ & \multicolumn{2}{c|}{$1.0\times10^8$} & \multicolumn{2}{c|}{$1.0\times10^8$}& \multicolumn{2}{c|}{$1.0\times10^8$} & \multicolumn{2}{c|}{$1.0\times10^8$}& \multicolumn{2}{c|}{$1.0\times10^8$} & \multicolumn{2}{c|}{$1.0\times10^8$}& \multicolumn{2}{c|}{$1.0\times10^8$} \\
\hline
&0&-&0&-&1&C&0&-&1&C&1&C&1&C\\
\cline{2-15}
$5\times10^4$ & \multicolumn{2}{c|}{$1.0\times10^8$} & \multicolumn{2}{c|}{$1.0\times10^8$}& \multicolumn{2}{c|}{$1.0\times10^8$} & \multicolumn{2}{c|}{$1.0\times10^8$}& \multicolumn{2}{c|}{$1.0\times10^8$} & \multicolumn{2}{c|}{$1.0\times10^8$}& \multicolumn{2}{c|}{$1.0\times10^8$} \\
\hline
&0&-&0&-&0&-&0&-&0&-&0&-&1&C\\
\cline{2-15}
$10^4$ & \multicolumn{2}{c|}{$1.0\times10^8$} & \multicolumn{2}{c|}{$1.0\times10^8$}& \multicolumn{2}{c|}{$1.0\times10^8$} & \multicolumn{2}{c|}{$1.0\times10^8$}& \multicolumn{2}{c|}{$1.0\times10^8$} & \multicolumn{2}{c|}{$1.0\times10^8$}& \multicolumn{2}{c|}{$1.0\times10^8$} \\
\hline
\end{tabular}
\end{minipage}
\end{table*}

\section{NUMERICAL RESULTS}

\subsection{Dependence on gas density and BH density}
The results of ten BHs systems are summarized in Table \ref{res_t1} and Table \ref{res_t2}. 
Table \ref{res_t1} is those for the case of $M_\mathrm{BH}=30 ~M_\odot$ and Table \ref{res_t2} 
is for $M_\mathrm{BH}=10^4 ~M_\odot$.
The column is the initial extension of the BH distribution ($r_\mathrm{typ}$) and
the corresponding BH density ($\rho_{\rm BH}$), while
the row is the assumed gas density ($n_\mathrm{gas}$) in the system. 
$N_m$ is the number of the merged BHs (nine means all BHs merged into one). 
The 'type' shows the merger mechanism, which is classified in the next section. 
Also, the termination time of simulations, $t_\mathrm{fin}$, is shown. 

In the case of $M_\mathrm{BH}=30 ~M_\odot$ (Table \ref{res_t1}), 
we find that if $n_{\rm gas}$ is in the range from $5\times10^6$
to $10^8 ~{\rm cm}^{-3}$, then all the BHs merge into one massive BH over six orders of BH density
($\rho_{\rm BH} = 72 - 7.2 \times 10^{7} ~M_\odot {\rm pc^{-3}}$).
Also, if the gas density is higher than $n_\mathrm{gas}= 5\times10^5 ~\mathrm{cm}^{-3}$, 
the BH merger occurs once at least in $100 ~\mathrm{Myr}$. 
In the recent numerical simulations on first stars (Greif et al. 2011; Susa 2013; Susa et al. 2014), 
the density of BH remnants is expected to be $\rho_{\rm BH} \approx 10^{7} ~M_\odot {\rm pc^{-3}}$
and the gas density is to be in the range of $n_{\rm gas} \approx 10^7-10^8 ~{\rm cm}^{-3}$.
Hence, the present numerical results imply that all the BH remnants from first stars are likely to merge into one BH. 

In the case of $M_\mathrm{BH}=10^4 ~M_\odot$ (Table \ref{res_t2}), 
we find that if $n_{\rm gas}$ is in the range from $5\times10^6$
to $10^9 {\rm cm}^{-3}$, then all the BHs merge into one massive BH over five orders of BH density
($\rho_{\rm BH} = 2.4 \times 10^{2} - 2.4 \times 10^{7} ~M_\odot {\rm pc^{-3}}$).
Interestingly, in the gas denser than $5\times10^6 ~{\rm cm}^{-3}$,  multiple BHs can merge into one,
even if the typical separation is greater than several pc. This result implies that the BH merger
is able to contribute significantly to the formation of SMBHs at high redshifts. 
Similar to the case of $M_\mathrm{BH}=30 ~M_\odot$, 
if the gas density is higher than $n_\mathrm{gas}= 5\times10^5 ~\mathrm{cm}^{-3}$, 
the BH merger occurs once at least. 

In both cases of BH mass, if the gas density is very high and simultaneously  
the BH density is very low (upper left in Table \ref{res_t1} and Table \ref{res_t2}), 
no merger occurs.
This is due to the fact that the deep gravitational potential by gas enhances the circular 
velocities of BHs and therefore a self-gravitating system of BHs is hardly able to form. 

As shown in terms of the types of merger in Table \ref{res_t1} and Table \ref{res_t2}, the merger mechanism 
changes systematically with the gas density ($n_\mathrm{gas}$) and the BH density ($\rho_\mathrm{BH}$).
Details of the merger mechanism are described in the following.

\subsection{Types of merger mechanism}

Here, we scrutinize the merger mechanism and classify the mechanism.
When a BH binary merges due to the dynamical friction by gas, the separation 
of the same pair of BHs monotonically shrinks, while the three-body interaction
between a close BH binary and an intruding BH often replaces one BH in the binary by
the intruder and therefore the separation of the closest pair changes violently. 
In Figs. \ref{30_1}-\ref{104_01}, we show the separation of the closest pair 
within all BHs as a function of time, 
where the colors of lines change at every event of the BH merger. 

Fig. \ref{30_1} shows the case of low-mass BHs ($M_\mathrm{BH}=30 ~M_\odot$).
The initial extension of the BH distribution is $r_\mathrm{typ}=1.0 ~\mathrm{pc}$.
The upper panel corresponds to higher gas density, and the lower panel to lower gas density. 
As seen in the upper panel, the apocenter and pericenter distances of the closest pair  
decay smoothly, eventually resulting in the merger. 
Such smooth decay indicates that the gas dynamical friction drives the merger.
On the hand, in the lower panel of Fig. \ref{30_1}, the orbit is highly eccentric, since the apocenter and pericenter distances 
are largely different. 
The distances sometimes oscillate violently. 
Such strong variance is thought to be caused by the three-body interaction, 
since the simultaneous interaction with a 4th or 5th BH seems quite rare from a viewpoint of probability. 
Actually, we have confirmed that the contribution of a 4th or 5th BH is little.
Thus, we call such interaction the three-body interaction. 
After the apocenter decays to less than $\sim 10^{-8} ~\mathrm{pc}$, the pair separation lessens rapidly
due to the effect of the GW emission, and eventually the BH binary merges.

We classify the merger mechanism according to the manner of decay just
before the GW promotes the merger.
Here, the merger mechanism is categorized into three types: 
gas-drag-driven merger (type A), three-body-driven merger (type C), and interplay-driven merger (type B). 
In type A, the dynamical friction by gas effectively 
decays the orbit and then the GW drives the merger.
The type A is seen in the cases with higher gas density and lower BH density, as expected. 
The examples of type A are shown in the top panels of Figs. \ref{30_1},\ref{30_01} and \ref{104_1}, and Fig. \ref{104_10}.
In both types B and C, the strong disturbance of the orbit is induced by the three-body interaction during the first merger. 
But in type B, after the first few mergers by three-body interaction,
the orbit decays slowly for long time through the gas dynamical friction. 
The examples of type B are seen in the middle panels of Figs. \ref{30_01} and \ref{104_1}, and Fig. \ref{104_01}.
In these results, after the mergers driven by three-body interaction, the orbit evolution starts from a larger separation
than initial typical separation ($r_\mathrm{typ}$). Such increase of separation 
is due to the slingshot mechanism when an intruder interacts with the binary. 
This causes a negative effect on the merger timescale. 
In other words, the three-body interaction may play a positive and negative role for the merger. 
\footnote{In the present study, we assume the uniform gas. But, in a real galaxy, the gas distribution 
is likely to become diffuse in outer regions. If the velocity of the kicked BH is high enough,
the BH cannot fall back but may escape.}
In type C, the strong disturbance of the orbit continues until the final merger.
The three-body interaction of BHs solely transfers the angular momentum, eventually
allowing the merger via the GW radiation.
Type C is seen in the cases of high BH density. 
The examples of type C are shown in the bottom panels of Figs. \ref{30_01} and \ref{104_1}. 

As shown in Tables 1 and 2, in the cases of lower BH density (higher $r_\mathrm{typ}$),
type A is seen in a broad range of gas density. 
In the cases of higher BH density, type B is dominant in higher gas density environments. 
On the other hand, type C is common in lower gas density. 
But, if the gas density is very low, no merger occurs even in the same density of BHs. 
This implies that the gas dynamical friction plays a non-negligible role even for type C.

\begin{figure*}
\includegraphics[width=120mm]{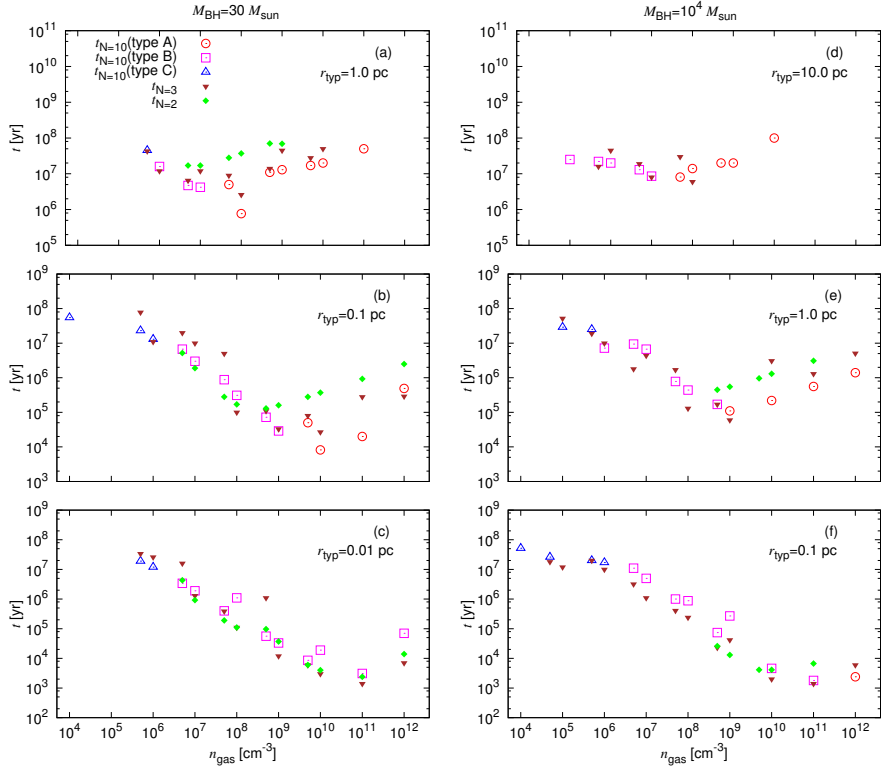}
 \vspace{12mm}
 \caption{Merger time in ten-BH systems ($t_{N=10}$) as a function of gas number density $n_\mathrm{gas}$. 
The left panels are the cases with low-mass BHs ($M_\mathrm{BH}=30 ~M_\odot$),
 and right panels are those with high-mass BHs ($M_\mathrm{BH}=10^4 ~M_\odot$). 
The top, middle and bottom panels show the results for lower, intermediate, and higher BH density, respectively. 
Red, pink, and blue open symbols represent the results of the types A, B, and C mergers, respectively. 
Green- and brown-filled symbols are the results of two-BH ($t_{N=2}$) and three-BH ($t_{N=3}$) systems, respectively.}
\label{23t}
\end{figure*}

\begin{figure*}
\includegraphics[width=120mm]{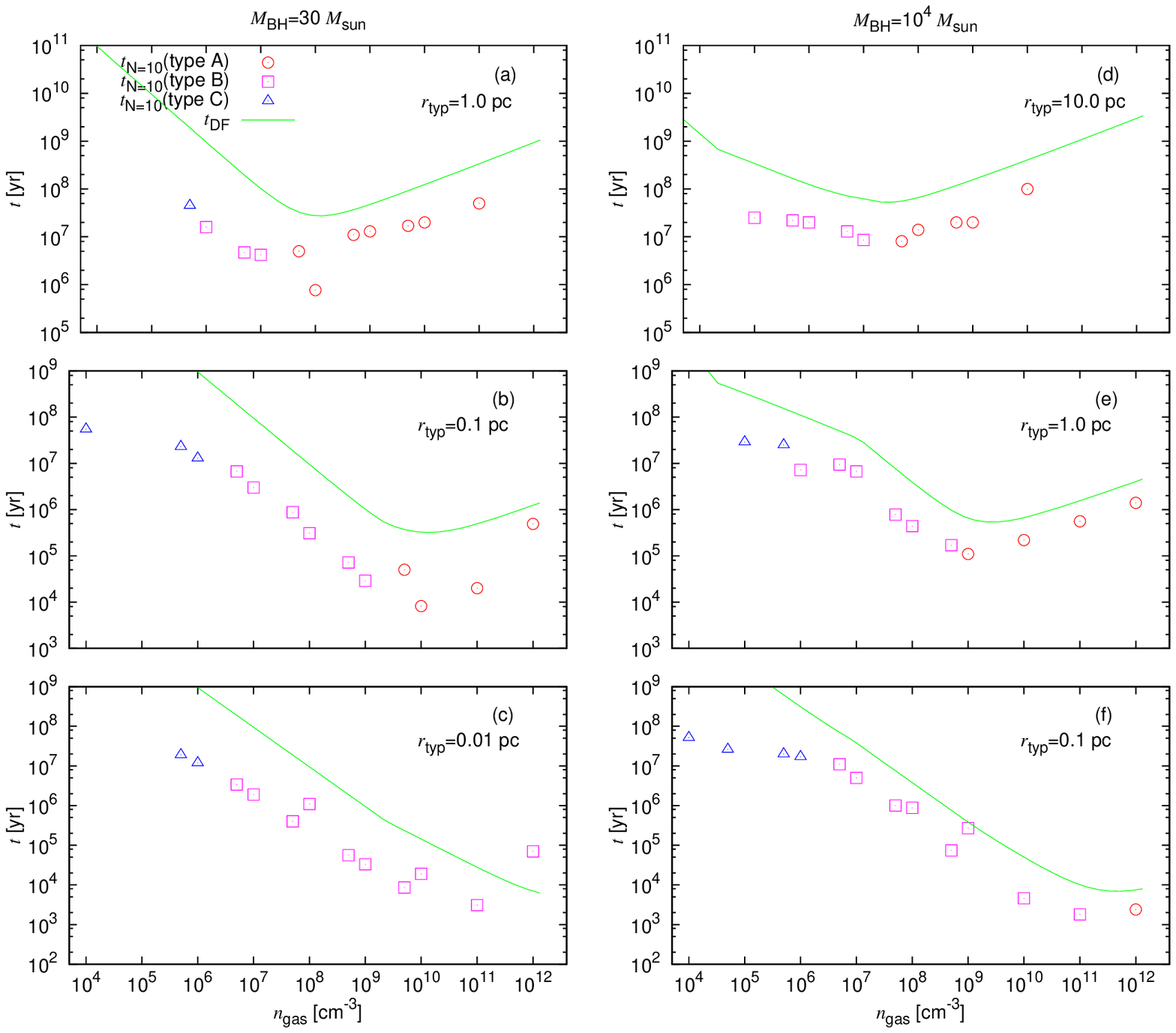}
 \vspace{12mm}
 \caption{
Same as Fig. \ref{23t}, but the analytic estimates of gas dynamical friction timescale, $t_\mathrm{DF}$, 
are plotted by green lines, instead of the results of two-BH and three-BH systems. 
}
\label{th10t}
\end{figure*}

\subsection{BH number dependence}

To clarify the physics of BH merger, we perform reference simulations 
with the different number of BHs.
In Fig. \ref{23t}, we compare the resultant merger time in ten-BH systems ($t_{N=10}$)
to that in three-BH ($t_{N=3}$)  or two-BH systems ($t_{N=2}$) . 
Each panel corresponds to a different set of parameters.
Red, pink, and blue open symbols respectively represent the types A, B, and C mergers in ten-BH systems.
Brown- and green-filled symbols show the averaged merger time in the runs of 
three- and two- BH systems, respectively. 
Note that the averaged merger time ($t_\mathrm{fin}$) has the variance of about a factor of two 
that comes from the initial setup of random number. 

In the cases of two BHs, we give only circular velocity initially. 
Since there is no three-body interaction, the merger of two-BH systems is always driven by the
gas dynamical friction. On the other hand, in the systems of three BHs, 
three-body interaction as well as the gas dynamical friction can work to induce the merger. 
We can see that the merger timescales in type C accord with those in three-BH systems.
In other words, no merger occurs in two-BH systems in the parameters of type C.
Thus, the classification of type C in which the merger is driven by the three-body interaction
is justified. 


On the other hand, in type A and type B, both two-BH and three-BH systems
can merge. In the parameters of type A, the merger timescales in ten-BH systems are systematically 
shorter than those in three-BH systems. In the ten-BH systems, more BHs suffer from 
the gas dynamical friction and the angular momenta of BHs are extracted simultaneously. 
Therefore, the ten-BH systems can merge in shorter time on average. 
As shown in top left and top right panels, two-BH or three-BH systems cannot merge
in higher gas density environments, whereas ten BHs systems can merge. 
These results imply that the dynamical friction on ten BHs enhances the three-body interaction, 
eventually allowing the merger.
In the bottom right panel, the averaged merger time in ten-BH systems is slightly longer 
than that in three-BH systems, in the cases with lower gas density ($n_{\rm gas} <10^8~\mathrm{cm^{-3}}$).
This is expected by the negative feedback as discussed in section 3.2.


 \subsection{Effect of recoil}
 
In the above, we have not considered the effects of gravitational wave recoil.
The recoil velocity ranges from several ten km~s$^{-1}$ to several thousand km~s$^{-1}$ (Kesden et al. 2010).
Tanikawa and Umemura (2014) have shown that if the recoil velocity is lower than
the virial velocity in the system, multiple BHs can eventually merge. 
In the present simulations, the virial velocity is higher than a few hundred km~s$^{-1}$
for $n_\mathrm{gas} \ge 10^8~\mathrm{cm}^{-3}$ and $M_\mathrm{gas}\ge 10^6~M_\odot$ 
in the case of $M_\mathrm{BH}=30~M_\odot$, and 
for $n_\mathrm{gas} \ge 10^6~\mathrm{cm}^{-3}$ and $M_\mathrm{gas}\ge 10^7~M_\odot$ 
in the case of $M_\mathrm{BH}=10^4~M_\odot$.
Actually, we have imposed random velocities on BHs at a level of the virial velocity, 
and revealed the conditions under which all BHs can merge. 
Hence, the merger condition is expected not to change, if the recoil velocity is lower than the virial velocity.
To confirm this, we have simulated the several cases in which the recoil velocity is incorporated. 
As a result, we have found that if the recoil velocity is lower than the virial (escape) velocity, then the averaged merger time 
and the number of merged BHs are not changed. 
If the recoil velocity is higher than the escape velocity, then the merged BH escapes from the system. 
The actual magnitude of recoil velocity is determined by the alignment of spins of two MBHs 
(Schnittman \& Buonanno 2007; Kesden et al. 2010).
The spins tend to be aligned through the gas accretion onto BHs. 
Assuming the level of the recoil velocity for BHs with aligned spins, we have performed further simulations and
found that if the escape velocity of the system is higher than $\sim140 \mathrm{km/s}$, then 
the BH merger proceeds in a similar way to the simulations without the recoil. 


\section{CRITERION FOR GAS DRAG}

In Fig. \ref{th10t}, we compare the averaged merger time in ten BHs systems ($t_\mathrm{N=10}$)
to the analytic estimate of gas dynamical friction timescale ($t_\mathrm{DF}$). 
We assess the friction timescale assuming a binary BH, 
similar to those by Begelman et al. (1980) and Matsubayashi et al. (2007). 
Based on a test simulation for a BH binary, we postulate that the eccentricity of the binary 
becomes very high, when the BH gravity dominates the gas potential.
Then, $t_\mathrm{DF}$ is estimated as 
\begin{eqnarray}
	t_\mathrm{DF}&=&\int_{100~r_\mathrm{sch}}^{r_\mathrm{typ}}\frac{\tilde{t}_\mathrm{DF}(r)}{r} dr\\	
	\tilde{t}_\mathrm{DF}(r)& \equiv & \frac{v_\mathrm{circ}(r)}{\max \{a^\mathrm{gas}_\mathrm{DF}(r')\} } \dots r \leq r' \leq r_\mathrm{typ}\\
	\tilde{t}_\mathrm{DF}(r_\mathrm{typ})&\simeq& \frac{v_\mathrm{circ}^3}{4\pi G^2 M_\mathrm{BH} m_\mathrm{H} n_\mathrm{gas}}\\
			&\simeq&
\left \{
\begin{array}{ll}
	\frac{(4\pi m_\mathrm{H} n_\mathrm{gas})^{1/2}r_\mathrm{typ}^3}{3^{3/2}G^{1/2} M_{\rm BH}} ~~( M_\mathrm{gas,b} \gg M_\mathrm{BH} ) \\
	\frac{M_{\rm BH}^{1/2}}{4\pi G^{1/2} m_\mathrm{H} n_\mathrm{gas} r_\mathrm{typ}^{3/2}}  ~~( M_\mathrm{gas,b} \ll M_\mathrm{BH} ) \label{t_DF_typ}
\end{array}
\right. 
\end{eqnarray}
Here, $M_\mathrm{gas,b}$ is the gas mass within the binary orbit, and $v_\mathrm{circ}$ is the circular velocity
given as $v_\mathrm{circ}=[G(2M_\mathrm{gas,b}+M_\mathrm{BH})/2r]^{1/2}$.
In this estimate, we use the maximum value of $a^\mathrm{gas}_\mathrm{DF}$ in the range of $r \leq r' \leq r_\mathrm{typ}$.
Green curves in Fig. \ref{th10t} present the estimated timescale of gas dynamical friction. 
According to equation (\ref{t_DF_typ}), 
$t_\mathrm{DF}$ is proportional to $n_\mathrm{gas}^{1/2}$ in the limit of high gas density, 
whereas  $t_\mathrm{DF}$ is inversely proportional to $n_\mathrm{gas}$ in the limit of low gas density. 
Thus, a turning point (the minimum) emerges in each curve of $t_\mathrm{DF}$ due to this change of dependence. 
As seen in Fig. \ref{th10t}, the turning point is in accordance with the transition from 
type A (gas-drag-driven merger) to type B (interplay-driven merger), and also
the trend of merger time ($t_\mathrm{N=10}$) is in agreement with the analytic estimate
of gas dynamical friction timescale ($t_\mathrm{DF}$). 
The timescale becomes longer in the higher gas density due to the increase 
in the deepness of gravitational potential of gas. 
The longer timescale in the lower gas density stems from the strong gravitational potential of other BHs. 
In addition, 
we can recognize the difference that the analytic estimate is systematically longer than 
the merger time in the simulations. 
One reason for the systematic difference comes from dismissing the  GW
as well as the three-body interaction effect in the analytic estimate. 
Another reason is concerns 
the fact that the analytic timescale is based on the approximate values of initial positions 
in type B and type C. 

In Fig. \ref{th10t}, we see the dependence of the merger timescale on the BH density (or $r_\mathrm{typ}$).  
With the increasing BH density (decreasing $r_\mathrm{typ}$), the right side of the power law shifts lower.
This shift stems from the decrease in the gravitational potential of gas within $r_\mathrm{typ}$. 
This tendency is also seen in the numerical results of ten-BH systems. 
Thus, the rough tendency of the merger time in the type A region is explained  by the analytic timescale.
Therefore, we regard $t_\mathrm{DF}$ as an appropriate estimate
for the merger solely by the gas dynamical friction. 

As described above, type A is always seen in higher gas density 
than the turning point of the double power law of $t_\mathrm{DF}$. 
These results imply that the turning point provides the critical gas density, above which the merger is  
driven by the gas dynamical friction.  
The turning point is determined by the ratio of the gas mass within $r_\mathrm{typ}$, $M_\mathrm{gas}$, to the total BH mass, $\sum M_\mathrm{BH}$.
In the analytic estimate, the turning point appears around $M_\mathrm{gas}\simeq 1.4\times10^5 \sum M_\mathrm{BH}$ 
as the average of the each panel in Fig. \ref{th10t}. 
As for the numerical results, we can assess the turning points in the top and middle panels, 
since the turning points are not clear in the bottom panels. 
In the cases of $30 ~M_\odot$ BHs, the gas density at the turning points 
is $10^7~\mathrm{cm^{-3}}$ in upper left panel or $10^{10}~\mathrm{cm^{-3}}$ in middle left panels.
Then, the ratio of $M_\mathrm{gas}$ to $\sum M_\mathrm{BH}$ at the turning points is 
$1.4\times 10^5$ in both cases.  
In the cases of $10^4 ~M_\odot$ BHs, it is $10^7~\mathrm{cm^{-3}}$
in upper right panel, or  $10^9 ~\mathrm{cm^{-3}}$ in middle right panels. Hence,  
the $M_\mathrm{gas}$-to-$\sum M_\mathrm{BH}$ ratio is $4.2\times 10^5$ or $4.2\times 10^4$, respectively.
These results show that the critical gas density above which type A is dominant is given by the condition of  
$M_\mathrm{gas}/\sum M_\mathrm{BH} \approx 10^5$.
Based on the simulations on the formation of first objects (Greif et al. 2011; Susa et al. 2014), 
$M_\mathrm{gas}$ is likely to be lower than $10^5\sum M_\mathrm{BH}$. Thus, type B and type C 
are expected to be more common mechanism of BH merger.

\section{CONCLUSIONS}

In this paper, we have investigated the merger of multiple BHs systems in an early cosmic epoch,
incorporating the dynamical friction by gas. For the purpose, we have performed 
highly accurate numerical simulations taking into account such general relativistic effects as
the pericentre shift and gravitational wave emission. 
Consequently, we have found the following: 

\begin{enumerate}
\renewcommand{\labelenumi}{\arabic{enumi}.}

\item \noindent
Multiple BHs are able to merge into one BH within 100 Myr in a wide range of parameters. 
In the case of $M_\mathrm{BH}=30 ~M_\odot$, 
if $n_{\rm gas}$ is in the range from $5\times10^6$ to $10^8 ~{\rm cm}^{-3}$, 
then all the BHs can merge for the BH density of 
$\rho_{\rm BH} = 72 - 7.2 \times 10^{7} ~M_\odot {\rm pc^{-3}}$. 
In the case of $M_\mathrm{BH}=10^4 ~M_\odot$, 
if $n_{\rm gas}$ is in the range from $5\times10^6$ to $10^9 ~{\rm cm}^{-3}$, 
then all the BHs can merge for the BH density of 
$\rho_{\rm BH} = 2.4 \times 10^{2} - 2.4 \times 10^{7} ~M_\odot {\rm pc^{-3}}$.
Commonly, if the gas density is higher than $n_\mathrm{gas}= 5\times10^5 ~\mathrm{cm}^{-3}$, 
the BH merger occurs once at least in $100 ~\mathrm{Myr}$. 
\\

\item \noindent
The merger mechanism is classified into three types: 
gas-drag-driven merger (type A),  three-body-driven merger (type C), and 
interplay-driven merger (type B). A criterion between 
type B and type C is almost concordant with the results of two- or three-BH systems. 
\\

\item \noindent
Based on the argument of merger timescales, type A merger has turned out to occur,
if the gas mass within the initial BH orbit ($M_\mathrm{gas}$) is higher than $10^5 \sum M_\mathrm{BH}$,
where $\sum M_\mathrm{BH}$ is the total mass of BHs. 
\\

\item \noindent
Supposing the gas and BH density based on the recent numerical simulations on first stars 
(Greif et al. 2011; Umemura et al. 2012, Susa 2013; Susa et al. 2014), 
all the BH remnants from first stars are likely to merge into one BH through the type B or C 
mechanism. 
\\

\item \noindent
As for a primordial galaxy possessing massive BHs with $M_\mathrm{BH}=10^4 ~M_\odot$, 
it is suggested that in environments of gas density higher than $5\times10^6~\mathrm{cm}^{-3}$,  
multiple BHs can merge into one through the type B mechanism,
even if the typical separation is greater than several pc. This result implies that the BH merger
is able to contribute significantly to the formation of SMBHs at high redshifts. 

\end{enumerate}

A point we have not considered in this paper is the effect of mass accretion onto BHs. 
During the motion of BHs in gas, some gas can fall onto BHs through the Bondi-Hoyle-Littleton accretion. 
This effect may significantly affect the merger processes of BHs. 
This issue will be studied in a forthcoming paper.
Besides, in a primordial galaxy, the stellar dynamical friction may work cooperatively with the gas friction
to extract the angular momenta. Such synergism is also an issue to be explored in the future.

\section*{Acknowledgments}

Numerical computations and analyses were carried out on Cray XC30 and computers at Center for Computational Astrophysics, 
National Astronomical Observatory of Japan, respectively.



\bsp

\end{document}